# Aligner-induced tooth movements in three dimensions using clinical data of two patients

**Ignacio Filippon [1], Christine Tanner [1,2,*], Jeannette A. von Jackowski [1,3], Georg Schulz [1,2], Tino Töpper [1], and Bert Müller [1,3]**

[1] Biomaterials Science Center, Department of Biomedical Engineering, University of Basel, 4123 Allschwil, Switzerland; ignacio.filippon@unibas.ch (I.F.); christine.tanner@unibas.ch (C.T.); jeannette.vonjackowski@unibas.ch (J.J.); georg.schulz@unibas.ch (G.S.); tino.toepper@unibas.ch (T.T.); bert.mueller@unibas.ch (B.M.)

[2] Core Facility for Micro- and Nanotomography, Department of Biomedical Engineering, University of Basel, 4123 Allschwil, Switzerland

[3] Biomaterials Science Center, Department of Clinical Research, University of Basel, 4031 Basel, Switzerland

* Correspondence: christine.tanner@unibas.ch

**Abstract:** The effectiveness of a series of optically transparent aligners for orthodontic treatments depends on the anchoring of each tooth. In contrast with roots, the crowns' positions and orientations are measurable with intraoral scans, thus avoiding any X-ray dose. Exemplified by two patients, we demonstrate that three-dimensional crown movements could be determined with micrometer precision by registering weekly intraoral scans. The data show the movement and orientation changes of the individual crowns of the upper and lower jaws as the result of the forces generated by the series of aligners. During the first weeks, the canines and incisors were more affected than the premolars and molars. We detected an overall tooth movement of up to about 1 mm during a nine-week treatment. The data on these orthodontic treatments indicate the extent to which actual tooth movement lags behind the treatment plan as represented by the aligner shapes. The proposed procedure can not only be used to quantify the clinical outcome of the therapy, but also to improve future planning of orthodontic treatments for each specific patient. The study should be treated with caution because only two cases have been investigated and the approach should be applied to a reasonably large cohort to reach strong conclusions regarding efficiency and efficacy of the therapeutic approach.

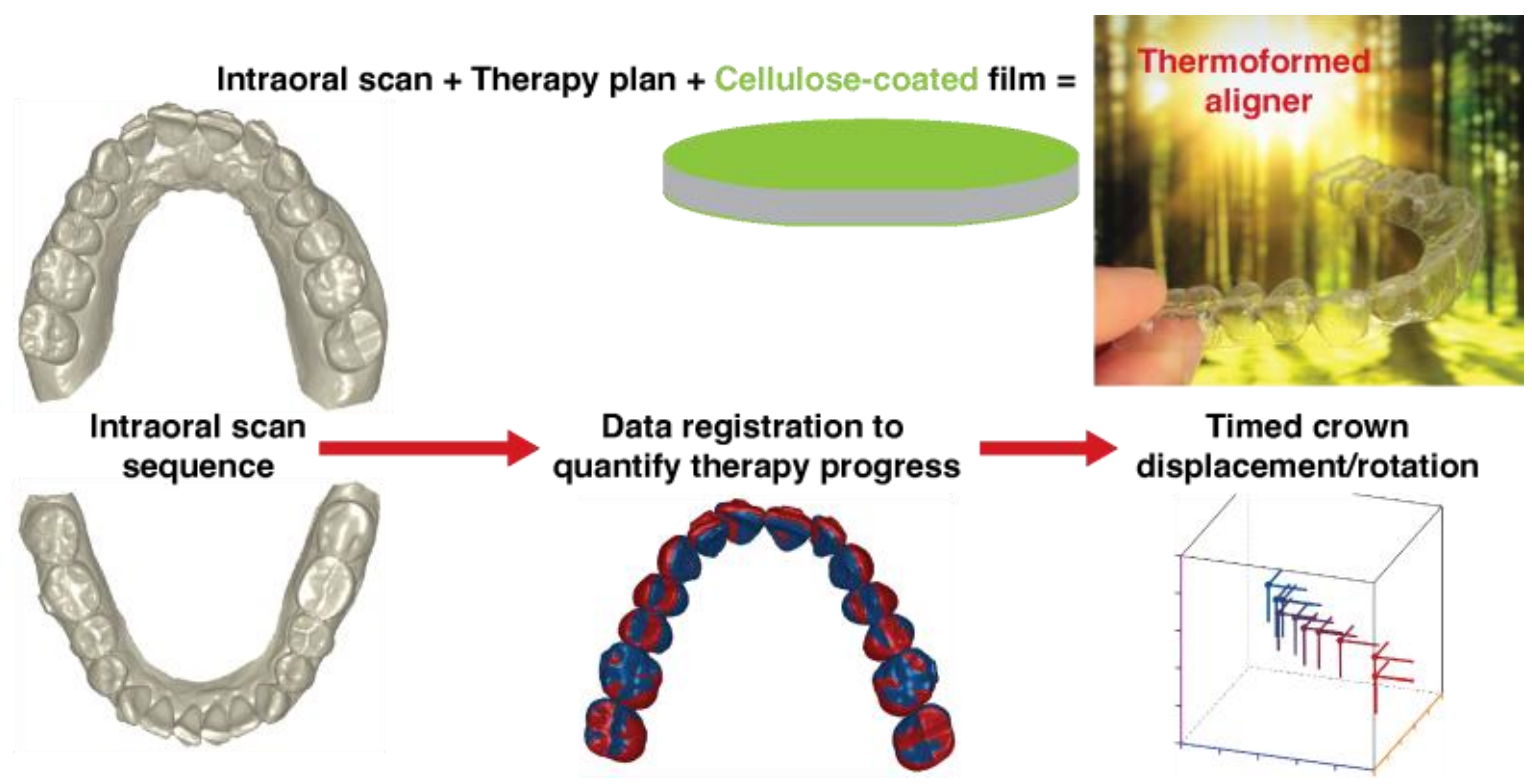

Graphical Abstract

## 1. Introduction

During the last two decades, optically transparent aligner therapy has become more and more popular, although orthodontic treatments still often rely on metal brackets [1,2]. This development is in large part because optically transparent aligners are perceived as an appealing option due to aesthetics and, more importantly, the possibility of removal for chewing food and tooth cleaning [3]. Furthermore, compared to traditional brackets, clear aligners provide a more stable course of treatment [4]. The digital workflow based on intraoral scanners has substantially facilitated aligner fabrication [5]. The performance of aligner therapy is continuously improving but not yet well quantified [6]. Some quantitative data, often combined with the clinical experience, are available [6]. The visualization of teeth, *i.e.* predictive, semiquantitative appearance, supports improved planning of aligner therapy [7]. The question arises: how can actual tooth movements and orientation changes be reasonably quantified?

An important parameter for successful therapy is the treatment duration [8]. There are well-known restrictions on the force limits to be applied [9]. Experts have generally agreed on two-week intervals for replacement with the successor aligner, see *e.g.* ref. [10]. It is hypothesized that a more frequent change might be beneficial [11]. Therefore, the present case study uses weekly aligner changes with a related intraoral scan during the first nine weeks of aligner therapy. Such an approach might be essential, because patients have negatively reflected on the treatment duration [12].

Conventional polymeric aligners give rise to microplastics and additives that are detected within the patient's body [13,14]. Therefore, the present study uses the Naturaligner™, a polyethylene terephthalate glycol-modified product with a cellulose-polymer coating and naturally derived plasticizers, which reduce the unwanted phenomenon of non-degradable microparticles release. Finally, the Naturaligner™ orthodontic therapy provides a reasonable fit to the teeth [15].

The tooth movement in orthodontic treatments has been described on the microscopic level [16]. Here, however, we use oral scans to assess movement because they are accessible to many clinics and do not use X rays. With these intraoral scanners, only the crowns can be quantified with limited spatial resolution, thus the proof of the related theories is restricted. With this study, we aim to propose a procedure for the quantification of the position changes of the individual crowns in the upper and lower jaws induced by aligner therapy.

We deal with the translational and rotational degrees of freedom (rotation, tip, inclination) for each individual crown, which is simply considered as a solid. Based on the weekly oral scans and their registration, the progress of tooth arrangement is quantified down to micrometer level. This series of oral scans of two patients acquired in the course of the therapy are combined and visualized by the movie in the supplementary materials. They are directly compared with the therapy plans to determine how much the actual orthodontic situation is behind the scheduled plans.

## 2. Materials and Methods

### *2.1. Intraoral scanner*

To select an intraoral scanner for the clinical study, the suitability and accuracy of the four available scanners: Primescan (Dentsply Sirona, Charlotte, NC, USA), TRIOS® 3 (3shape, Copenhagen, Denmark), iTero Element 5D Plus scanner (Align Technology, San Jose, CA, USA), and Medit i700 (Medit corp., Seongbuk-gu, South Korea) were compared following the published procedure [17]. As model served a maxillary full denture with three parallel hollow cylinders at tooth positions 17 (C1), 21 (C2), and 27 (C3), see ref. [17] and Figure 2(a). It was milled out of an industrially manufactured polyether ether ketone (PEEK) block (Denseo PEEK blank, Denseo GmbH, Aschaffenburg, Germany) on a five-axis computerized-numerical-control milling machine (SilaMill 5, vhf camfacture

AG, Ammerbuch, Germany, with a 3 μm repeat accuracy) as previously described in Section 2.1 of ref. [17]. The PEEK model was scanned three times by an experienced dentist (I.F.). The operator exactly followed the guidelines of the suppliers.

Microtomography data of the PEEK model were used as reference. We selected the forth dataset described in ref. [17] because of the high accuracy and its match to the CAD model. The three-dimensional data were acquired with a conventional tomography system, nanotom® m (phoenix|x-ray, GE Sensing & Inspection Technologies GmbH, Wunstorf, Germany), using an effective pixel length of 35 μm and averaging more than 120 voxels per reference center, which increases the accuracy to about 6 μm [17,18]. It works, because the nanotom® m is designed for sub-micrometer resolution imaging, which was confirmed by investigations with a Siemens star [19] and, therefore, provides the necessary mechanical stability.

The model surfaces obtained by the intraoral scanners were compared with the reference surface by the following approach. First, the crown-like features and the hollow cylinders of the reference were manually segmented using the software FreeCAD (version 0.19), see yellow and magenta objects in Figure 2(b), respectively. The center of each hollow cylinder was determined by the geometrical mean of its thousands of surface points. Second, the surfaces from the scans were manually pre-aligned. Third, the reference surface was rigidly registered to the scans using the iterative closest point method from MATLAB (release R2020b, The MathWorks, Inc., Natick, USA) with optimization parameters set as follows: random sampling of 50% of points, 70% inlier ratio, 500 maximum number of iterations, and 2.5 μm and 0.125 deg as convergence tolerance for translation and rotation, respectively. Fourth, for every point of the rigidly transformed reference surface, the closest point of the scanned surface was determined. Several points from the reference surface were often matched to the same point on the scanned surface, since the reference surface is finer sampled as the scanned meshes, i.e. 82 vs. 98 to 158 μm mean distance between points. This discretization error was removed by replacing many-to-one distances by their smallest distance. Fifth, the error distributions were summarized by the mean and standard deviation of all points of the reference mesh regions and of 100 randomly sampled points of the reference mesh regions. Two regions were evaluated, namely either the three hollow cylinders or the surface of the crown-like features. Statistical significance of differences in median values of the Euclidean distances was tested using the Mann-Whitney U-test. Statistical significance of differences in mean values of the log-Euclidean distances was tested using the two-sample t-test.

Finally, we decided to apply the intraoral scanner Primescan (Dentsply Sirona, Charlotte, NC, USA) for the case study, see below. The planning of the orthodontic therapy, the Naturaligner™ fabrication, and the determined translation and rotation of crowns were based on the scans recorded by the experienced dentist (I.F.). The weekly acquired scans are labeled according to the time points Week 0, *i.e.* the start of the therapy, to Week 9.

*2.2. Fabrication of optically transparent aligners with cellulose coating*

The company Bottmedical AG, Basel, Switzerland provided sets of aligner pairs fabricated from 550 μm-thin cellulose-coated polyethylene terephthalate glycol-modified films [15]. These films were thermoformed, because this approach leads to a reasonable fit [20] with an average thickness of 334 μm [15]. A heating temperature of 220 °C of a standard thermoforming unit Biostar® (SCHEU-DENTAL GmbH, Iserlohn, Germany) resulted in a film surface temperature of 190 °C. A pressure of 4.8 bar was applied to the films pressed towards a 3D-printed model prepared according to the therapy planning. After separation from the model, the aligner was trimmed to guarantee a suitable interface to the gingiva. The optically transparent aligners are sold under trade name Naturaligner™.

*2.3. Case study with two patients*

The two female patients had a dental class II and moderate to severe crowding for both jaws according to the Discrepancy Index of the American Board of Orthodontics (ABO), see below. Informed consent forms were signed by the patients.

The patients' teeth were prepared according to the following standard procedure: first, the teeth were cleaned with a fluoride-free paste with Pumice powder; second, the phosphoric acidic gel Ultra-Etch™ (Ultradent Products Inc., Köln, Germany) was applied for a period of 30 seconds followed by water rinsing and drying with air. Third, 3M™ Transbond™ XT was brushed on the treated enamel surfaces, aired, and light-cured for a duration of 20 seconds. Subsequently, the aligner attachments were glued using 3M™ Transbond™, before the excess composite has been gently removed, and the attachments were polished.

Right after each appointment for intraoral scanning, the patients were instructed how to change the aligners of the upper and lower jaws. This means the first change of each new aligner was done under supervision. In addition, the patients obtained an identical pair of aligners to replace them midweek. This approach was chosen to ensure that a potential deformation of the 550 μm-thin aligners during one week could be avoided and a moderate and constant force is maintained for inducing the desired tooth movements during the therapy.

The patients were advised to wear the aligners for a duration of at least 22 hours per day. For eating and teeth brushing, the aligners should be carefully removed. Patients were allowed to drink with aligners in place, considering the potential pigmentation of the devices. They were instructed to gently clean the aligners twice per day with a soft tooth brush.

While the case study with two patients has insufficient statistical power for providing conclusive clinical results, it could show that the proposed method is suitable for being applied to a reasonably large cohort.

*2.4. Selection of the global and local coordinate systems*

Figure 1 shows the intraoral scan data at Week 9 of Patient A and Patient B. Three anatomical landmarks at the two distobuccal cusps on the second molars for the upper and lower jaws as well as labels between the crowns of the central incisors were manually selected to determine the global coordinate system, see black dots and black arrows [21]. These landmarks define the global left and global extrusion directions, where the latter is indicated by the magenta-colored arrows on the individual crowns. The global anterior direction is selected orthogonal to these two axes.

Figure 1 also shows the coordinate systems for each crown using colored arrows. The extrusion direction, given by the magenta color, coincides with the global coordinate system. The mesial direction of each tooth is given by the vector to the neighboring crown and displayed by the orange-colored arrows. The blue-colored arrows indicate the buccal direction of each crown and are orthogonal to the orange- and magenta-colored arrows.

It should be noted that for Patient A one third molar and for Patient B the four third molars are missing, see Figure 1. Therefore, the third molars have not been included into the movement study. Patient A had an orthodontic treatment previously. During this treatment, the four first premolars were extracted and thus Figure 1 (a) shows only 24 crowns besides the third molars.

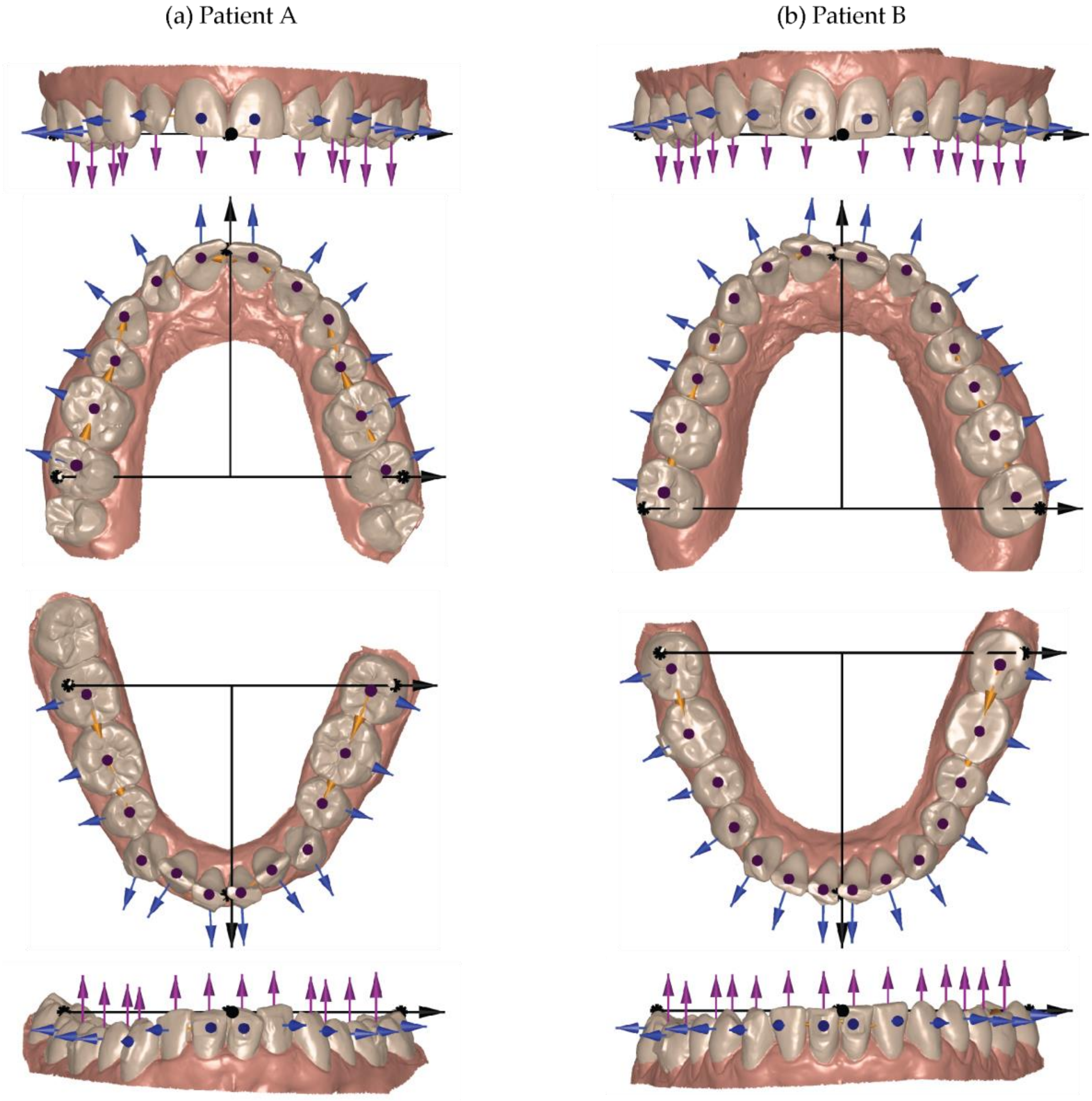

**Figure 1.** Based on the intraoral scans of the upper and lower jaws at Week 9, the black arrows/dots illustrate the selection of the global coordinate systems for the patient data, see left and right column, respectively. The orange-, magenta-, and blue-colored arrows are used to display the coordinate systems for the individual crowns in mesial, extrusion, and buccal directions, respectively.

### *2.5. Data evaluation*

The individual crowns at state Week 9 were semiautomatically segmented from the three-dimensional surface data of the intraoral scans for the upper and lower jaws by an experienced dentist (I.F.) using the commercially available software OnyxCeph³™ (Image Instruments GmbH, Chemnitz, Germany). The necessary interactions of the operator took less than 25 minutes per jaw.

The jaw surfaces were trimmed with manually defined polygons using FreeCAD (version 0.19) to reduce the surface to the teeth and the adjacent gum to approximately the mucogingival junction. The segmented crowns were fused to a single surface mesh and rigidly registered to the trimmed jaw surfaces of the oral scans acquired at the selected timepoints using the MATLAB iterative closed points function called *pcregistericp*. Starting from this rigid registration result, the surface of the crowns were non-rigidly registered to the other trimmed jaw surfaces using the MATLAB function *nricp* based on the implementation from *github.com/charlienash/nricp* [22]. This non-rigid motion field defined the orthodontic tooth motion. Translation and rotation per crown in the three-dimensional domain were determined by fitting a rigid transformation to the non-rigid motion field of the crown, after setting the center of rotation to the crown's center of mass. Note, this center of rotation serves to describe the rigid transformation of the crown in three-dimensional space, while the anatomical center of rotation of the tooth is unknown. The fitting was based on minimizing the least root mean square distance via the Kabsch algorithm [23] using the implementation from the MATLAB *fileexchange 25746-kabsch-algorithm*.

The teeth segmentations at Week 9 were propagated to the states at Weeks 0 to 8 based on the closest points after the non-rigid registration. Orthodontic tooth motion between Week 0 and all other states was determined by rigid and non-rigid registration as described above. Comparison to the treatment plan was similarly determined by rigid and non-rigid registration.

## 3. Results

### *3.1. Intraoral scanner selection*

Figure 2 (a) displays the surface of the microtomography data from the PEEK model used as reference. The segmented crown-like features in yellow and the segmented hollow cylinders in magenta were used to determine the accuracy of the four intraoral scanners available. In a first step, we checked whether the four scanners fully capture these features of the PEEK model. The images of the intraoral scanners in Figure 2 indicate that the crown-like features and the hollow cylinders are comprehensively represented. The scanned meshes had a mean distance between the points of 98, 146, 135, and 158 μm for the Sirona, Trios, iTero, and Medit devices, respectively. Thus, the data quality is reasonable. A more detailed quantitative analysis of the data is required to decide on the choice of the intraoral scanner for the present case study and related future clinical studies.

Table 1 provides the mean values and the related standard deviation of the error statistics calculated for the analyzed features. For the cylinder centers and both feature regions as well as for the entire data and the arbitrarily selected 100 points, the mean Euclidean distance is increasing from Sirona via Trios and iTero to Medit. The mean values of the Trios and iTero scanners are very similar.

The distributions from pooling three times 100 randomly selected points per scanner had a skew shape, as the Euclidean distance measure is always positive, see boxplots in Figure 3, first row. Taking the logarithm of the Euclidean distances, cf. boxplots in the second row of Figure 3, the distributions are almost symmetric. For both regions, *i.e.* the hollow cylinders and the crown-like features, the median and mean results from the Sirona scanner are significantly more accurate at the 0.001 level than the median and mean values of the other three intraoral scanners, respectively.

**Table 1.** Mean values ± standard deviation, in μm, of the Euclidean distances in three dimensions between the centers of the magenta-colored hollow cylinders and meshes for all and 100 randomly sampled points (in brackets) of hollow cylinders and crown-like features.

| Scanner | Cylinder centers | Hollow cylinders | Crown-like features |
|---|---|---|---|
| **Sirona** | **14 ± 8** | **55 ± 35 (56 ± 38)** | **50 ± 25 (48 ± 21)** |
| **Trios** | **24 ± 15** | **63 ± 26 (64 ± 28)** | **58 ± 24 (58 ± 24)** |
| **iTero** | **25 ± 15** | **65 ± 28 (66 ± 30)** | **60 ± 26 (58 ± 24)** |
| **Medit** | **43 ± 22** | **91 ± 90 (90 ± 91)** | **70 ± 49 (69 ± 40)** |

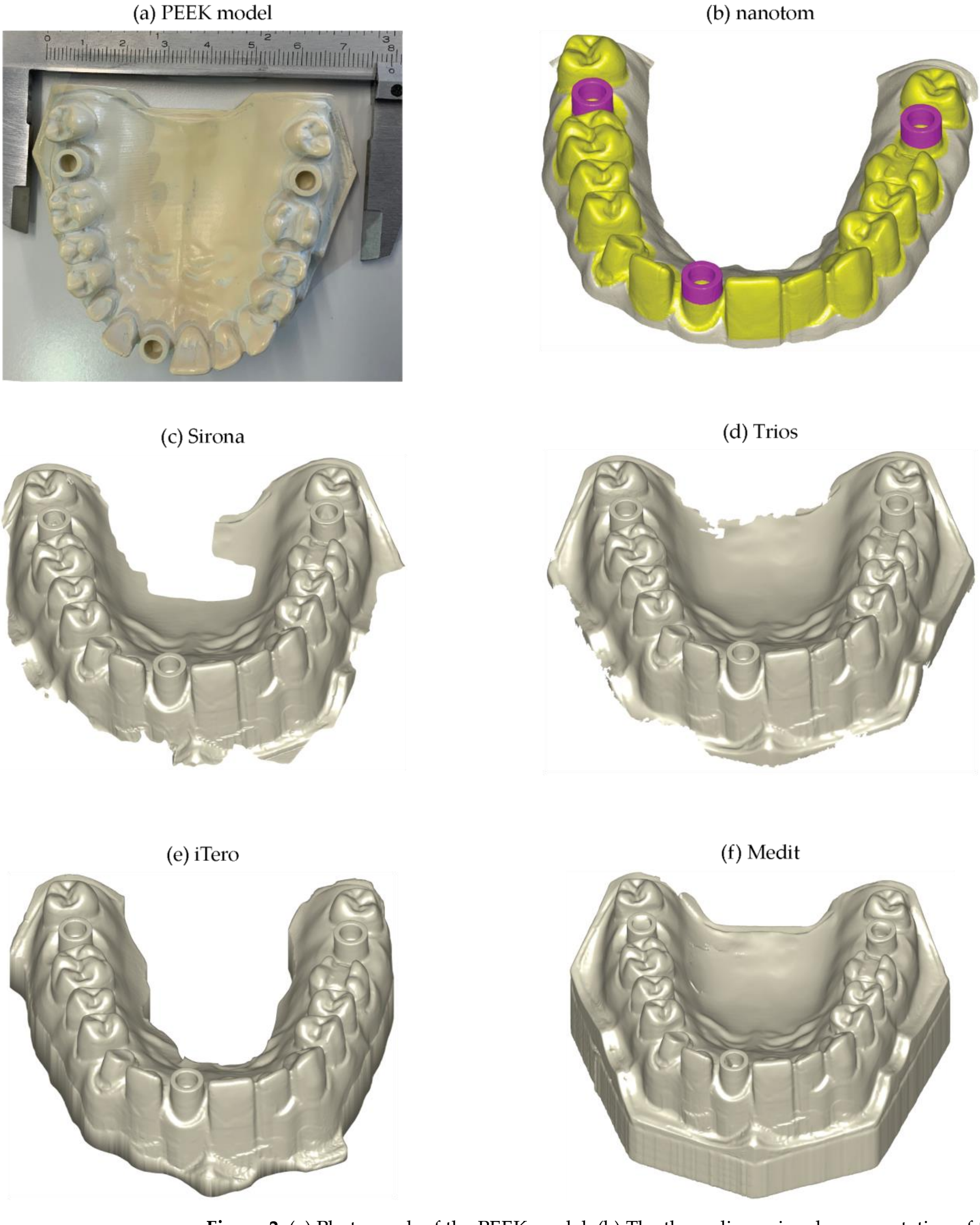

**Figure 2.** (a) Photograph of the PEEK model. (b) The three-dimensional representation of the reference microtomography data (model) show crown-like features in yellow and hollow cylinders in magenta colors. The data from the intraoral scanners (c) Sirona, (d) Trios, (e) iTero, and (f) Medit were manually aligned to the model and obtained as triplicates. The four devices allowed capturing the crown-like features and the hollow cylinders.

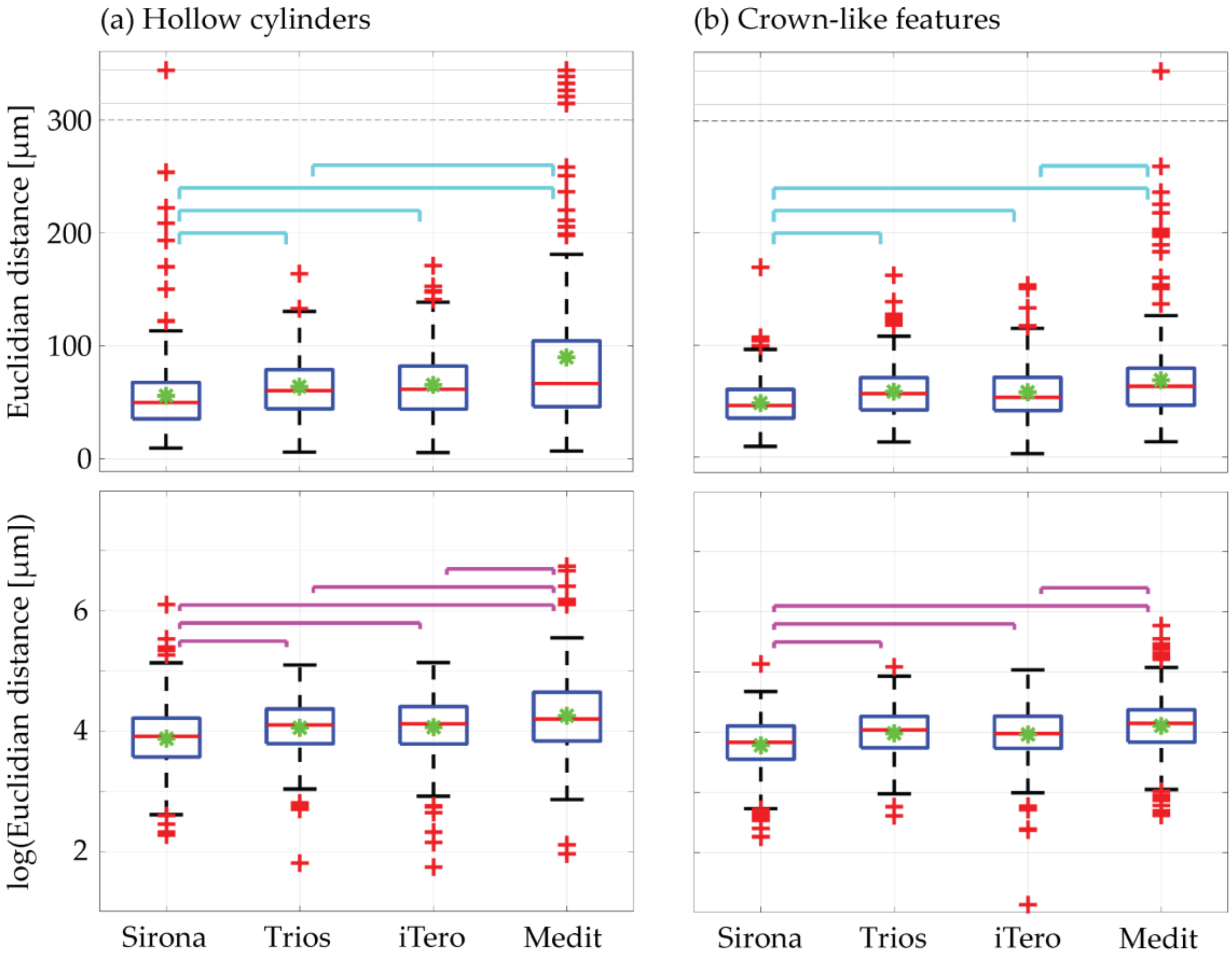

**Figure 3.** Distribution of three times 100 sampled Euclidean distances for (a) the hollow cylinders and (b) the crown-like features. Statistical significance at the 0.001 level is marked by cyan lines for median (Mann-Whitney U-test) and magenta lines for mean (t-test). The results for the hollow cylinders coincide with data previously published [17]. The data from the crown-like features show similar results. The intraoral scanner from Sirona is the best choice. Hence, it was selected for the present case study.

The results for the hollow cylinders can be directly compared to the previously reported data [17]. In this study, median Euclidean distances for the cylinders were smaller than previously reported, namely 61 vs. 79 µm for Trios and 67 vs. 98 µm for Medit. As before, median results were more accurate for the Trios scanner than for the Medit scanner.

The intraoral scans of Sirona were statistically significantly more accurate than the scans from the Trios, the iTero, and the Medit scanners. Consequently, the Sirona scanner was the best choice for the present case study with two patients.

*3.2. Tooth movement induced by a series of optically transparent aligners*

Figure 4 displays the arrangement of the crowns for the upper and lower jaws of the two patients at the therapy start (Week 0). These arrangements of crowns are given in quadruplicate to visualize the progress of therapy by colors from blue via green to yellow for the treatment duration of two, four, six, and eight weeks. The dark blue color labels the static situation: no tooth displacement with respect to the start of therapy. The gradual color change to yellow, see color bar of Figure 4, corresponds to the tooth movement magnitude of up to 1 mm. For Patient A, the gradual tooth movement during the selected therapy period is obvious. As expected, the tooth movement of canines and incisors is much larger than the one of the molars for both maxilla and mandible. Nevertheless, a slight movement of the first molar is observed. For Patient B, the tooth movement is also distinctly visible in the anterior region. The movement of the molars is delayed in treatment time and becomes only evident after a therapy duration of eight weeks.

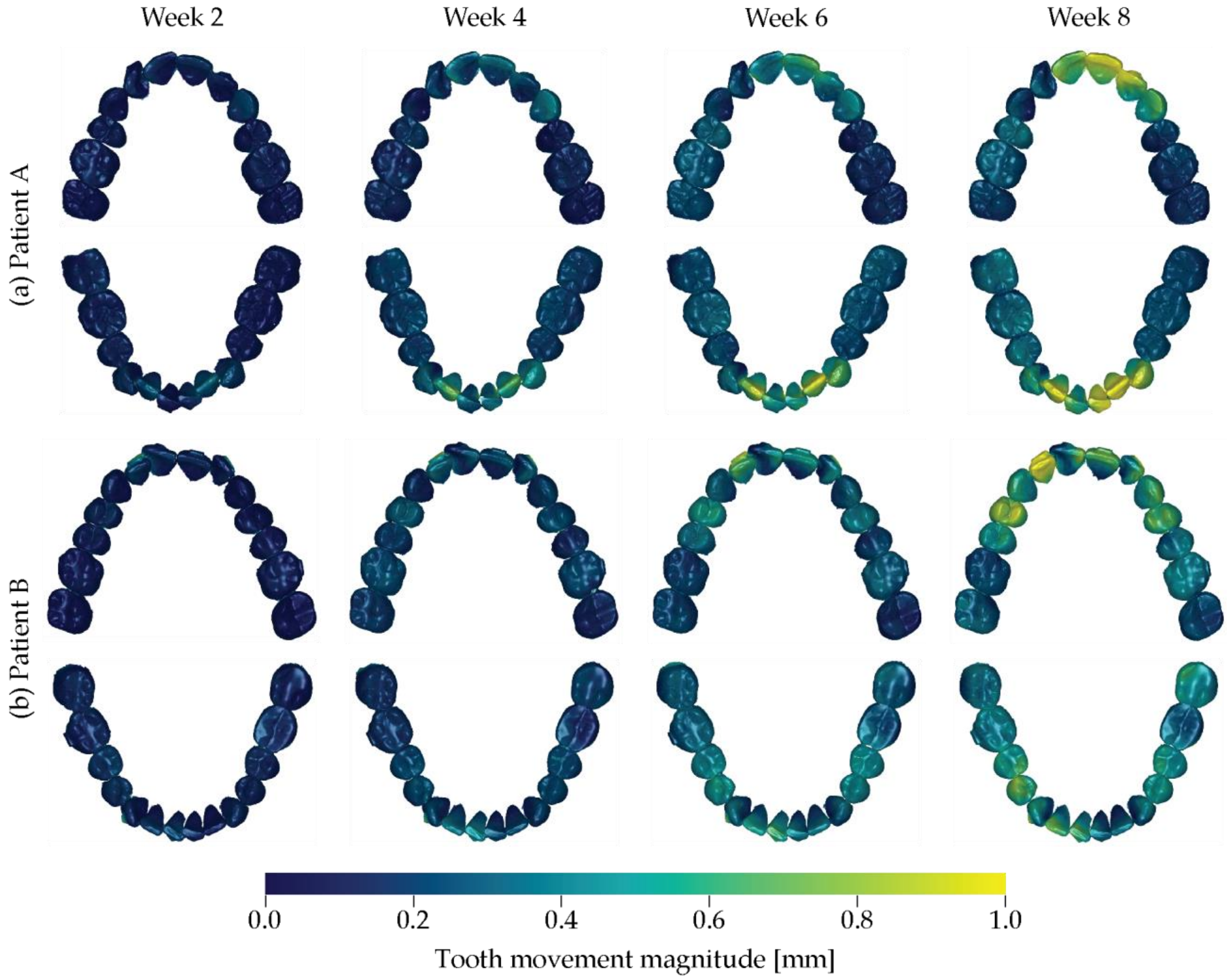

**Figure 4.** The arrangement of the crowns of the upper and lower jaws of Patient A and of Patient B is represented for the starting of the treatment, *i.e.* Week 0. Their color corresponds to the magnitude of orthodontic tooth movement ranging from blue via green to yellow, which is given for the selected treatment times Week 2, Week 4, Week 6, and Week 8, always with respect to Week 0. The movie in the supplementary materials shows the treatment progress.

### *3.3. Intrusion-extrusion crown displacements*

The intrusion and extrusion of the individual crowns is visualized for the two patient treatments in Figure 5. The intrusion-extrusion crown displacements consistently increase with the therapy time for the selected period.

The anterior regions of Patient A show intrusion of up to 0.4 mm. The first upper right molar and the second lower right molar also show intrusion. The extrusion of the first lower right molar is consistent with the intrusion of its antagonist. The first upper left premolar and first molar on this side show an intrusion of about 0.2 mm.

Patient B shows a more symmetric behavior than Patient A. The tooth displacement in intrusion and extrusion directions was achieved. The extrusion of the lower second molar was successful and improved the frontal deep bite. An especially strong intrusion could be observed in the lower second left molar, which is consistent with the small extrusion of its antagonist in the upper left region for leveling the arch.

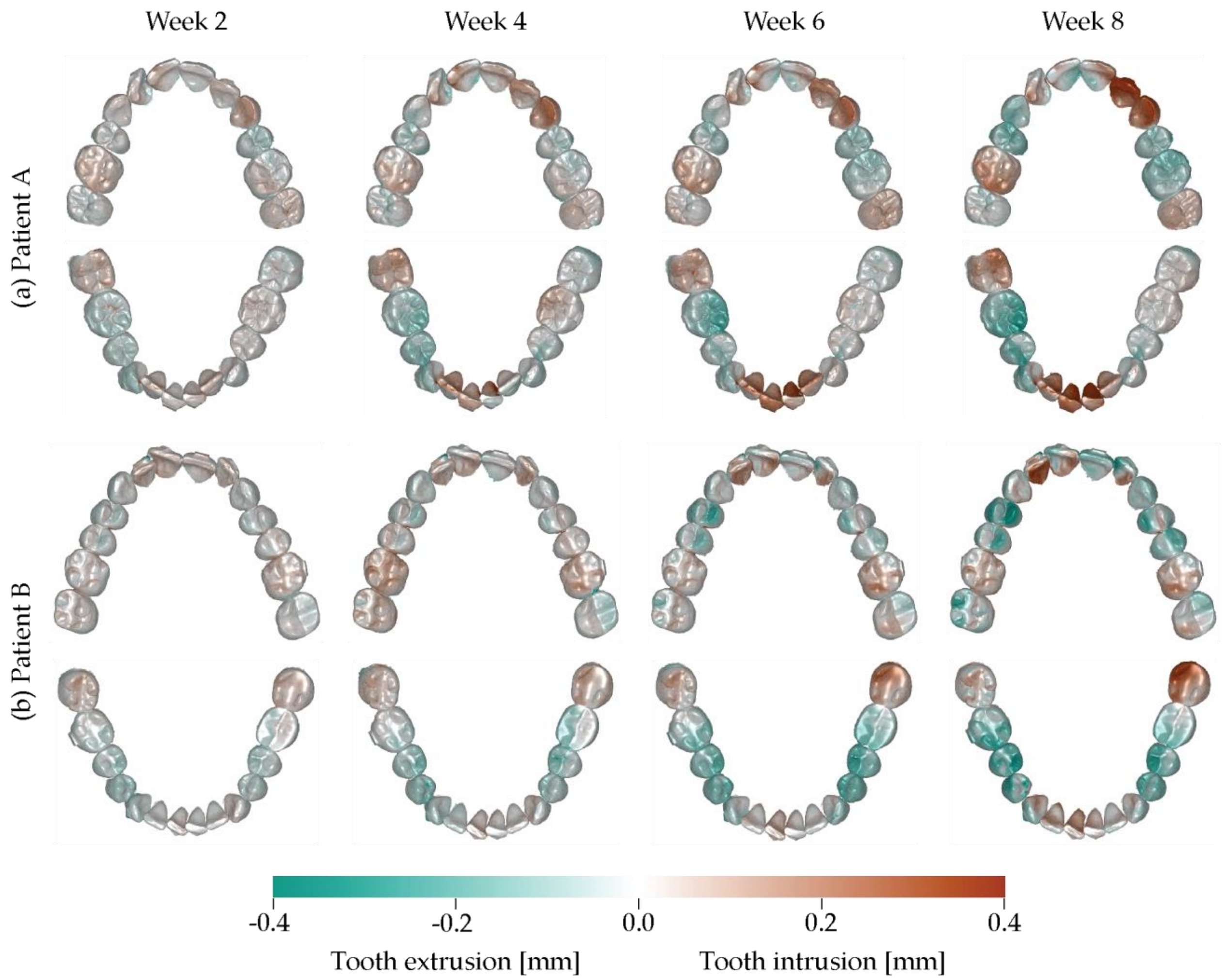

**Figure 5.** The tooth displacement in the extrusion and intrusion directions is labeled according to the color bar from mint to reddish color. Looking at the individual jaws, one frequently recognizes opposite coloring for the opposite quadrants. The reference is always the start of the therapy, *i.e.* Week 0. The movie in the supplementary materials indicates the temporal changes in tooth displacement.

### *3.3. Displacement of individual teeth*

The individual crowns were regarded as solids that do not change their shape during the therapy time of nine weeks. Therefore, three translational and three rotational parameters per crown are sufficient to quantitatively describe their movement, as evident in Figures 6 and 7 for Patient A and Patient B, respectively.

The surface representations in Figures 6 and 7 show the displacement of the crown to be considered from the therapy start in dark blue and to the situation after nine-week aligner treatment in dark red. The translation and rotation in time steps of one week using colors from blue to red are visualized in three-dimensional space by the cubes with a volume of $(1\ mm)^3$.

In general, major changes are observed for the incisors and canines, whereas the posterior regions are hardly affected. The incisors of Patient A showed relatively large translational and rotational changes. The premolars and molars exhibited much less changes, especially in the posterior region were molars only underwent minor rotations and minimal intrusions. Patient B presented a similar picture. Some intrusion of the canines in both jaws could be detected. In comparison with Patient A, we could identify larger rotational changes in the premolars than in molars, more prominent in the mandible.

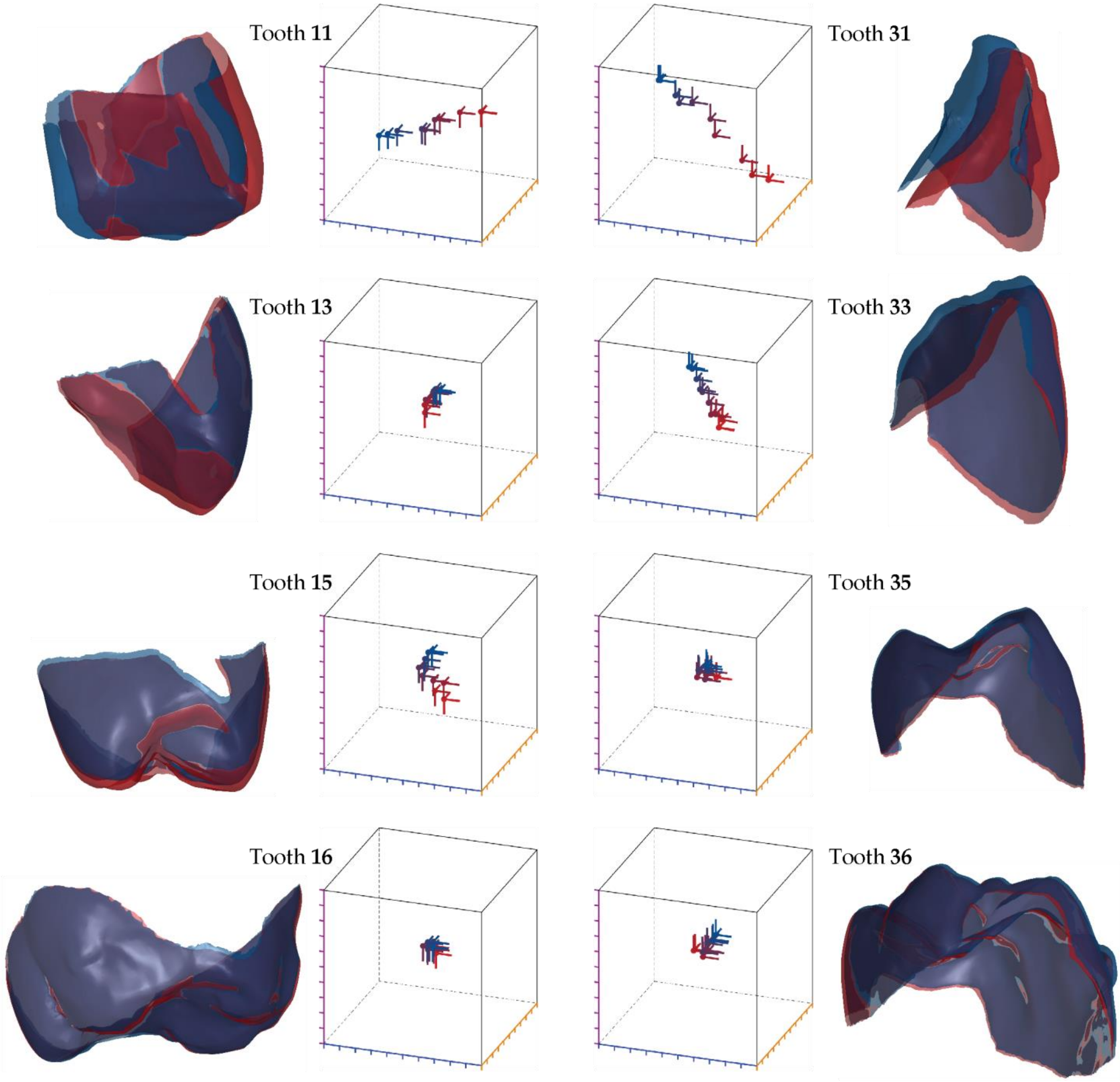

**Figure 6.** Movement of selected teeth of Patient A on the left for the maxilla and on the right the opposite teeth of the mandible. The renderings are the crown positions at the start of the treatment given in dark blue color and after nine-week treatment in dark red color. Translation and rotation for each treatment step, which corresponds to exactly one week, represented by blue to red color are displayed in the local coordinate system of the start state: intrusion-extrusion direction – magenta scale, distal-mesial direction – orange scale, lingual-buccal direction – blue scale. The boxes cover a volume of $(1\ \text{mm})^3$.

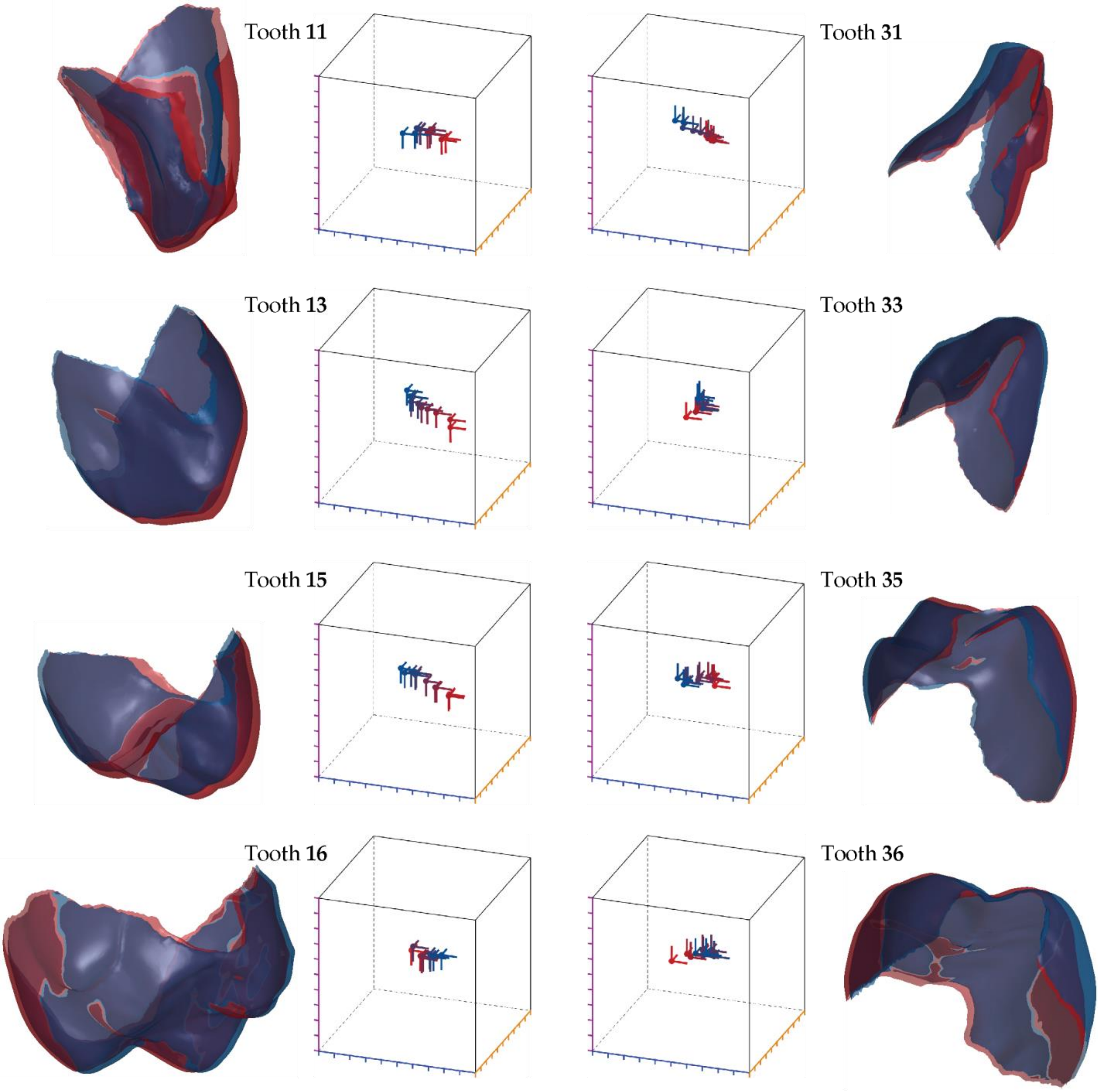

**Figure 7.** Movement of selected teeth of Patient B on the left for the maxilla and on the right the opposite teeth of the mandible. The renderings are the crown positions at the start of the treatment given in dark blue color and after nine-week treatment in dark red color. Translation and rotation for each treatment step, which corresponds to exactly one week, represented by blue to red color are displayed in the local coordinate system of the start state: intrusion-extrusion direction – magenta scale, distal-mesial direction – orange scale, lingual-buccal direction – blue scale. The boxes cover a volume of $(1\ mm)^3$.

### *3.4. Rate of tooth movement*

The results presented in Figure 8 focused on one of the objectives for this study, *i.e.* the quantification of tooth movement with time. Figure 8 depicts the rate of motion per week for each individual crown in the context of its neighborhood. It was observed that the neighboring teeth often moved together and that the motion rate generally increased at a later stage of therapy, for example for Patient A after a duration of five weeks.

For Patient A, we observed that in the maxilla the incisors and left canine, *i.e.* Tooth **11**, Tooth **21**, and Tooth **22**, firmly moved between Week 6 and Week 8. The rate of movement for the other teeth was rather constant. In the mandible, incisors but also canines and premolars showed an increased rate of movement, but more intermittent and not constant throughout the selected therapy steps.

For Patient B, the lateral incisor of the maxilla, Tooth **12**, showed a similarly high rate of movement that in Patient A. Less pronounced, it was also found for the other incisors, canines and premolars, see Week 8. In the right posterior of the mandible, see Tooth **37**, Tooth **36**, Tooth **35**, Tooth **34** and Tooth **33**, we detected a high rate of movement during the final stage of therapy, *i.e.* within Week 9.

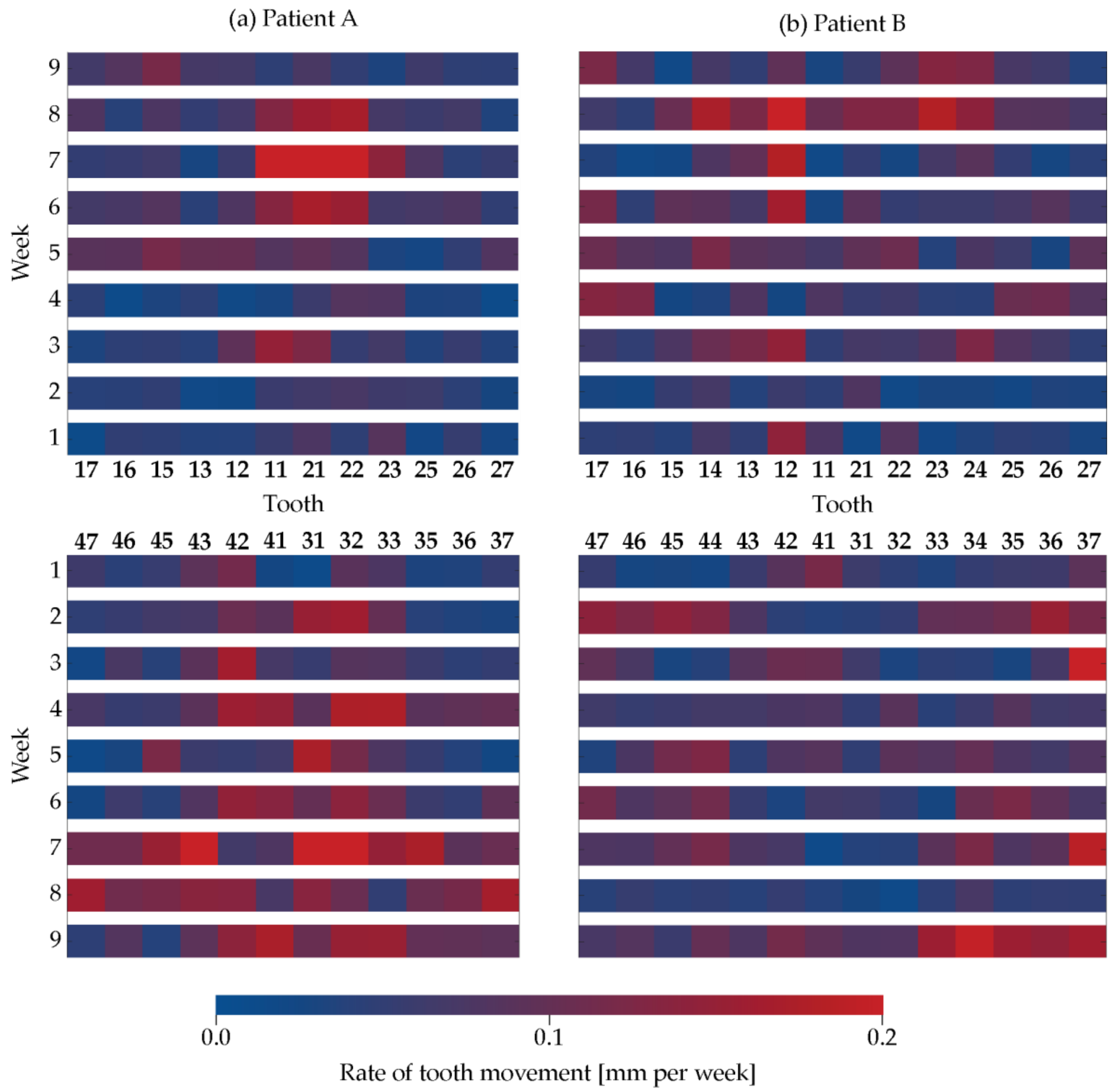

**Figure 8.** The rate of movement of the individual crowns is color-coded according to the bar and displayed for the aligner treatment duration according to the weekly oral scans: (a) upper and lower jaws of Patient A and (b) upper and lower jaws of Patient B.

*3.4. Tooth displacement difference to the therapy plan*

Figure 9 shows the difference between the therapy plan and the achieved aligner treatment after two, four, six, and eight weeks. We found differences of up to 1 mm represented by the red color. The planned buccal rotation for Tooth **17** and Tooth **27** was not achieved within the eight-week aligner treatment for both patients. In contrast, the incisors changed their position closely to the therapy plan.

For Patient A, the clinical outcome followed the therapy plan during the first five weeks. Subsequently, we found discrepancies, which were prominent in the fourth quadrant. We also identified a larger disagreement for the first than for the second quadrant.

For Patient B the main difference between planning and achieved treatment related to the molar of the maxilla with a lack of tip, *i.e.* rotation around the mesial-distal axis. In addition, one can recognize a substantial discrepancy for the premolars in the first quadrant of the maxilla at Week 4 and 6. At Week 8, however, it becomes closer to the virtual plan. In the mandible, the differences were less prominent with values below 0.6 mm.

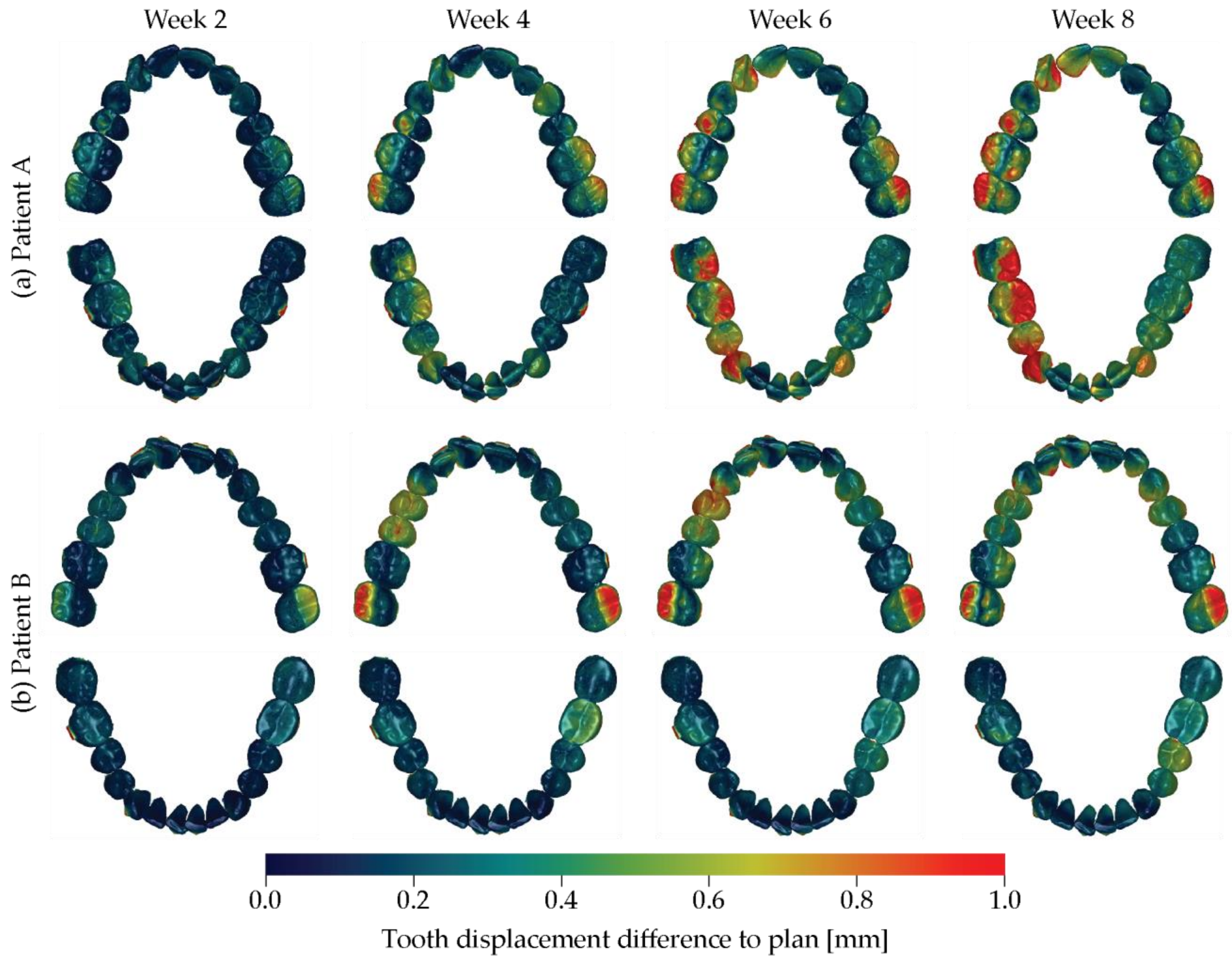

**Figure 9.** The difference of the actual tooth position and the one of the treatment plan is color-coded from blue via green to red, see color bar, for Patient A on the top and Patient B at the bottom for the four treatment durations indicated. The reference space is always the starting point, *i.e.* Week 0. The movie in the supplementary materials represents the progress of the actual treatment.

## 4. Discussion

Optically transparent aligners have been applied for more than two decades. Their application, however, has been only partially successful, as stated by many recent reviews, see *e.g.* [24-26]. Even very recent original work demonstrates the limitation of polymeric aligner treatments, see *e.g.* [27]. Therefore, orthodontists are sometimes doubtful about the predictability of tooth movements caused by aligner therapy. Aligner treatments in moderate to severe cases take more time than conventional brace therapies, since the tooth movement rate, related to the applied forces, is lower for the aligners that are usually made from polymers such as polyurethane and ethylene vinyl acetate. The force generation, however, can be adapted by the aligner thickness, see *e.g.* [28]. For effective bodily tooth movement, rather thick and single-layered rigid polymers are preferred [28]. Multi-layered aligners generally exert lower forces compared to their single-layered counterparts [28]. Therefore, we can reasonably expect comparably low force transmission from the 550 µm-thin cellulose-coated polyethylene terephthalate glycol-modified films of the Naturaligner™. That is why in the present study we have accelerated the treatment plan and applied two aligners per week to compensate for aging, as described in ref. [29].

To validate the treatment success of aligner therapy, intraoral scans at the start and the terminate state were collected and compared. Refinement scans, as done in the present case study, are often included [30]. The present study is based on weekly scans during the first nine weeks to precisely determine the orientation and location changes of the individual crowns. It should be emphasized that the intraoral scans only capture the crowns. Orthodontists are likely to change their aligner treatment plan when given three-dimensional information on the position of a patient's crowns, root, and bone coverage [31]. Within the current case study, we intentionally abstained from X-ray imaging of the upper and lower jaws, since the radiation dose for three-dimensional X-ray images would be inappropriate. This lack of information is a limitation of the study. A very recent study shows the benefits of using cone-beam computed tomography in evaluating clear aligners to conventional brace treatments [32].

Although automatic tools for segmenting the individual crowns from intraoral scans reached a certain maturity, manual adjustments are needed for clinical use [33]. For the current study, we thus selected the software OnyxCeph³ ™. Even though this software contains a module for the automatic crown segmentation, we have used the module for trimming the cast bases to ensure the necessary quality of segmenting the individual crowns at Week 9. In this fashion, each crown was separated from the arch. For this purpose, the points that define the crown boundaries were manually set. Depending on the crown size and shape, the number of points selected ranged from 10 to 15. The time needed to select these points decreased with experience. Thanks to the quality of the segmentation at Week 9 and the automatic non-rigid registration, the crowns of the earlier time points have been segmented promptly.

Adult patients show interest in fast aligner treatments [34]. Fialho *et al.* [35] compared aligner treatments of mature adults with seven- and 14-day exchange rates, which show the limitations of accelerated protocols. Nevertheless, we have decided to use a weekly exchange. Furthermore, we provided two identical aligners per week and advised the patients for a change after 3.5 days to compensate for the Naturaligner™ wear.

Posterior intrusion, mesiodistal translation, or torque movement are possible contraindications for the use of aligners, see *e.g.* ref. [36] and references therein. We could not confirm some of these findings within this case study. The discrepancy may be due to the limited accuracy of previous approaches. For Patient A, we detected the largest intrusion movements in the anterior region of the incisors of the lower jaw, which fits the planned correction of the frontal deep bite. The chronological sequence of individual crown displacement and rotation was consistent with the leveling of the arches and the Curve of Spee for both jaws. The real situation certainly lagged behind the planning. The direction and magnitude of movement of opposite teeth was consistent with the desired intrusion or extrusion. Crown rotations and especially extrusions of the molars were only partially achieved.

This communication concentrates on the first nine weeks of treatment, although it continued to 16 and 20 weeks, respectively. This continuation was necessary to reach full satisfaction by the patients and the attending dentist. It should be noted that during the aligner treatment the upper incisors of Patient B are mainly displaced, see movie in the supplementary materials, whereas the subsequent treatment induces tilting, see movie in the supplementary materials.

## 5. Conclusions

Treatments using optically transparent aligners are currently popular to displace and rotate the teeth in a desired fashion, as the numerously cited paper of Robertson *et al.*

clearly indicate [37]. We could demonstrate for two cases that the Naturaligner™ enabled a fast and desired orthodontic therapy. The proposed analysis, comprising of intraoral scans, manual crown segmentation at a selected stage, and automatic non-rigid registration of the entire sequence, enables the quantification of the displacement and rotation of the individual teeth over treatment time. It should become the basis for the analysis of a statistically relevant cohort with the goal to further improve aligner-based orthodontic therapies. Especially important, the proposed methodology will allow for the precise quantification of the difference between the actual arrangement of the crowns and the force- and torque-generating shape of the optically transparent aligner used. Nonetheless, the present study has two clear limitations. First, three-dimensional X-ray images have been regarded as inappropriate, so that the study only relies on the optical images of the crowns. Second, the informative value is restricted to only two cases and has to be extended to give strong conclusion about efficiency and efficacy.

**Author Contributions:** Conceptualization, J.J., T.T., and B.M.; methodology, C.T. and B.M.; software, C.T.; validation, C.T.; investigation, I.F. and G.S.; resources, I.F., J.J., G.S., T.T., and B.M.; data curation, I.F. and C.T.; writing—original draft preparation, I.F., C.T. and B.M.; writing—review and editing, I.F., C.T., J.J., G.S., T.T. and B.M.; visualization, C.T. and B.M.; supervision, B.M.; project administration, B.M.; funding acquisition, T.T. and B.M. All authors have read and agreed to the published version of the manuscript.

**Funding:** We thank Bottmedical AG, Basel, Switzerland for the study materials (Naturaligner™).

**Informed Consent Statement:** Written informed consent was obtained from the two patients involved in this case study.

**Data Availability Statement:** The data from the intraoral scans are available at the Zenodo repository DOI 11280343/zenodo.11280343 and the software scripts can be accessed via the Github repository toothMovementAnalysis.

**Acknowledgments:** The authors thank Zahnärzte am Kreis AG, 8754 Netstal, Switzerland for the support with the OnyxCeph³ ™ software and acknowledge the valuable technical discussions with Bekim Osmani, Basel, Switzerland. Griffin Rodgers kindly improved the readability of the manuscript.

**Conflicts of Interest:** I.F., C.T. and G.S. declare no conflicts of interest. J.J., T.T., and B.M. are founding members of Bottmedical AG, Basel, Switzerland. T.T. is board member, and B.M. is scientific advisor of Bottmedical AG, Basel, Switzerland.